\documentclass[twocolumn]{aastex631}

\newcommand\teff{T$_{eff}$}

\shorttitle{A brown dwarf in the JWST ERS Abell 2744 field}
\shortauthors{Nonino et al.}

\usepackage{hyperref}  
\hypersetup{colorlinks=true,linkcolor=[rgb]{1.,0.2,0.2},citecolor=[rgb]{0.1,0.4,1.},filecolor=[rgb]{0.7,0.2,0.2},urlcolor=[rgb]{0.7,0.2,0.2}}

\usepackage{color}
\definecolor{blue}{rgb}{0., 0., 1}

\def\micron{\ifmmode \mu\mathrm{m} \else $\mu$m\fi}
\def\msun{\ifmmode \mathrm{M}_{\odot} \else M$_{\odot}$\fi}
\def\mstar{\ifmmode \mathrm{M}_{\star} \else M$_{\star}$\fi}

\begin{document}

\title{Early results from GLASS-JWST. XIII. A faint, distant, and cold brown dwarf \footnote{ Based on data collected with JWST}}

\correspondingauthor{Mario Nonino}
\email{mario.nonino@inaf.it}

\author[0000-0001-6342-9662]{Mario Nonino }
\affiliation{INAF-Trieste Astronomical Observatory, Via Bazzoni 2, I-34124, Trieste, Italy}

\author[0000-0002-3254-9044]{Karl Glazebrook}
\affiliation{Centre for Astrophysics and Supercomputing, Swinburne University of Technology, PO Box 218, Hawthorn, VIC 3122, Australia}

\author[0000-0002-6523-9536]{Adam J. Burgasser}
\affiliation{Department of Physics, University of California San Diego, La Jolla, CA 92093, USA;}

\author[0000-0003-4067-9196]{Gianluca Polenta}
\affiliation{Space Science Data Center, Italian Space Agency, via del Politecnico, I-00133, Roma, Italy}

\author[0000-0002-8512-1404]{Takahiro Morishita}
\affiliation{IPAC, California Institute of Technology, MC 314-6, 1200 E. California Boulevard, Pasadena, CA 91125, USA}

\author[0000-0003-1287-9801]{Marius Lepinzan}
\affiliation{INAF-Trieste Astronomical Observatory, Via Bazzoni 2, I-34124, Trieste, Italy}

\author[0000-0001-9875-8263]{Marco Castellano}
\affiliation{INAF Osservatorio Astronomico di Roma, Via Frascati 33, I-00078 Monteporzio Catone, Rome, Italy}

\author[0000-0003-3820-2823]{Adriano Fontana}
\affiliation{INAF Osservatorio Astronomico di Roma, Via Frascati 33, I-00078 Monteporzio Catone, Rome, Italy}

\author[0000-0001-6870-8900]{Emiliano Merlin}
\affiliation{INAF Osservatorio Astronomico di Roma, Via Frascati 33, I-00078 Monteporzio Catone, Rome, Italy}

\author{Andrea Bonchi}
\affiliation{Space Science Data Center, Italian Space Agency, via del Politecnico, I-00133, Roma, Italy}

\author[0000-0002-7409-8114]{Diego Paris}
\affiliation{INAF Osservatorio Astronomico di Roma, Via Frascati 33, I-00078 Monteporzio Catone, Rome, Italy}


\author[0000-0002-8460-0390]{Tommaso Treu}
\affiliation{Department of Physics and Astronomy, University of California, Los Angeles, 430 Portola Plaza, Los Angeles, CA 90095, USA}

\author[0000-0003-0980-1499]{Benedetta Vulcani}
\affiliation{INAF Osservatorio Astronomico di Padova, Vicolo dell'Osservatorio 5, I-35122 Padova, Italy}

\author[0000-0002-9373-3865]{Xin Wang}
\affil{Infrared Processing and Analysis Center, Caltech, 1200 E. California Boulevard, Pasadena, CA 91125, USA}

\author[0000-0002-9334-8705]{Paola Santini}
\affiliation{INAF Osservatorio Astronomico di Roma, Via Frascati 33, I-00078 Monteporzio Catone, Rome, Italy}

\author[0000-0002-5057-135X]{E.~Vanzella}
\affiliation{INAF -- OAS, Osservatorio di Astrofisica e Scienza dello Spazio di Bologna, via Gobetti 93/3, I-40129 Bologna, Italy}

\author[0000-0003-2804-0648 ]{Themiya Nanayakkara}
\affiliation{Centre for Astrophysics and Supercomputing, Swinburne University of Technology, PO Box 218, Hawthorn, VIC 3122, Australia}

\author[0000-0001-9261-7849]{Amata Mercurio}
\affiliation{INAF -- Osservatorio Astronomico di Capodimonte, Via Moiariello 16, I-80131 Napoli, Italy}

\author[0000-0002-6813-0632]{P.~Rosati}
\affiliation{Dipartimento di Fisica e Scienze della Terra, Universit\`a degli Studi di Ferrara, Via Saragat 1, I-44122 Ferrara, Italy}

\author[0000-0002-5926-7143]{Claudio Grillo}
\affiliation{Dipartimento di Fisica, Università degli Studi di Milano, Via Celoria 16, I-20133 Milano, Italy}
\affiliation{INAF - IASF Milano, via A. Corti 12, I-20133 Milano, Italy}

\author[0000-0001-5984-0395]{M.~Bradac}
\affiliation{
University of Ljubljana, Department of Mathematics and Physics, Jadranska ulica 19, SI-1000 Ljubljana, Slovenia}
\affiliation{
Department of Physics and Astronomy, University of California Davis, 1 Shields Avenue, Davis, CA 95616, USA}



\begin{abstract}
{ We present the serendipitous discovery of a late T-type brown dwarf candidate in JWST NIRCam  observations of the Early Release Science Abell 2744 parallel field. The discovery was enabled by the
sensitivity of JWST at 4~{\micron} wavelengths and the panchromatic 0.9--4.5~{\micron} coverage 
of the spectral energy distribution.
The unresolved point source has magnitudes F115W = 27.95$\pm$0.15 and F444W = 25.84$\pm$0.01 (AB), and its F115W$-$F444W and F356W$-$F444W colors match those expected for other known T dwarfs.
We can exclude it as a reddened background star, high redshift quasar, or  a very high redshift galaxy.  Comparison with stellar atmospheric models indicates a temperature of {\teff} $\approx$ 650~K and surface gravity $\log{g}$ $\approx$ 5.25, implying a mass of 0.03~M$_{\odot}$ and age of 5~Gyr. We estimate the distance of this candidate to be 570--720~pc in a direction perpendicular to the Galactic plane, making it a likely thick disk or halo brown dwarf. These observations underscore the power of JWST to probe the very low-mass end of the substellar
mass function in the Galactic thick disk and halo.} 
\end{abstract}


\section{introduction} \label{sec:intro}

A significant fraction of Milky Way stars near the Sun are  cool brown dwarfs,
nonfusing stars that have masses M$\leq$0.07~M$_{\odot}$, effective temperatures (in the field) of {\teff} $\leq$ 2000~K, and are classified as late-{\em L, T,} and {\em Y} dwarfs \citep{1962AJ.....67S.579K,2000ARA&A..38..337C,kirkpatrick2005}. 
These objects are intrinsically faint and emit primarily at infrared wavelengths. 
Hence, wide-field imaging surveys of brown dwarfs are largely limited to the immediate solar neighborhood ({\em d $\leq 100$} pc), and few metal-poor thick disk and halo brown dwarfs have been identified to date \citep{2003ApJ...592.1186B,2019MNRAS.486.1260Z,2020ApJ...898...77S}. Deep pencil-beam surveys, conducted primarily with the {\it Hubble Space Telescope} ({\it HST}) WFC3 instrument, have extended brown dwarf detection to J $\approx 25.5$ (AB), reaching kpc scales \citep{ryan2011, masters2012, aganze2022}.
However, these surveys are restricted to wavelengths $<2$~{\micron}, limiting their sensitivity to the reddest and coldest brown dwarfs.
{\it The James Webb Space Telescope} ({\it JWST}) represents a major step forward in the detection of cool and distant brown dwarfs, with 
imaging and spectroscopy extending to  $\sim 28.3\micron$\  and providing orders of magnitude greater sensitivity than {\it Spitzer}, particularly in the $3-5\micron$\ wavelength range where cold brown dwarf spectral energy distributions peak.

The availability of the deep {\it JWST} NIRCam observations in a deep extragalactic
field have provided the first opportunity to study low-temperature brown dwarfs in the Galactic thick disk and halo. In this paper, we report the first detection of such an object. 
In Section 2, we summarize the {\it JWST} observations. 
In Section 3, we present the identification of the brown dwarf candidate, and compare its colors and magnitudes to theoretical models to estimate its physical properties. We also rule out potential contaminant sources. 
In Section 4, we present our conclusions.
In this paper, we use the AB magnitude system unless otherwise indicated.

\section{observations}

The {\it JWST} Director’s Discretionary Early Release Science Program ERS 1324 (PI T.\ Treu; hereafter GLASS-ERS)
targets the massive galaxy cluster Abell 2744 with NIRSPEC and NIRISS, and
simultaneously images a parallel field 3'--8'  away from the cluster with NIRCam
in seven broadband filters: F090W, F115W, F150W, F200W, F277W, F356W, and F444W.
These observations span the full 1--5 \micron\ range with unprecedented depth and spatial resolution. 
Details of the GLASS-ERS data acquisition and observing strategy can be found in \citet{Treu2022}.
Details of the image reduction, stacking, and catalog creation can be found in \citet{merlin2022}. 

\begin{figure*}
\centering
    
\includegraphics[scale=0.4,angle=0]{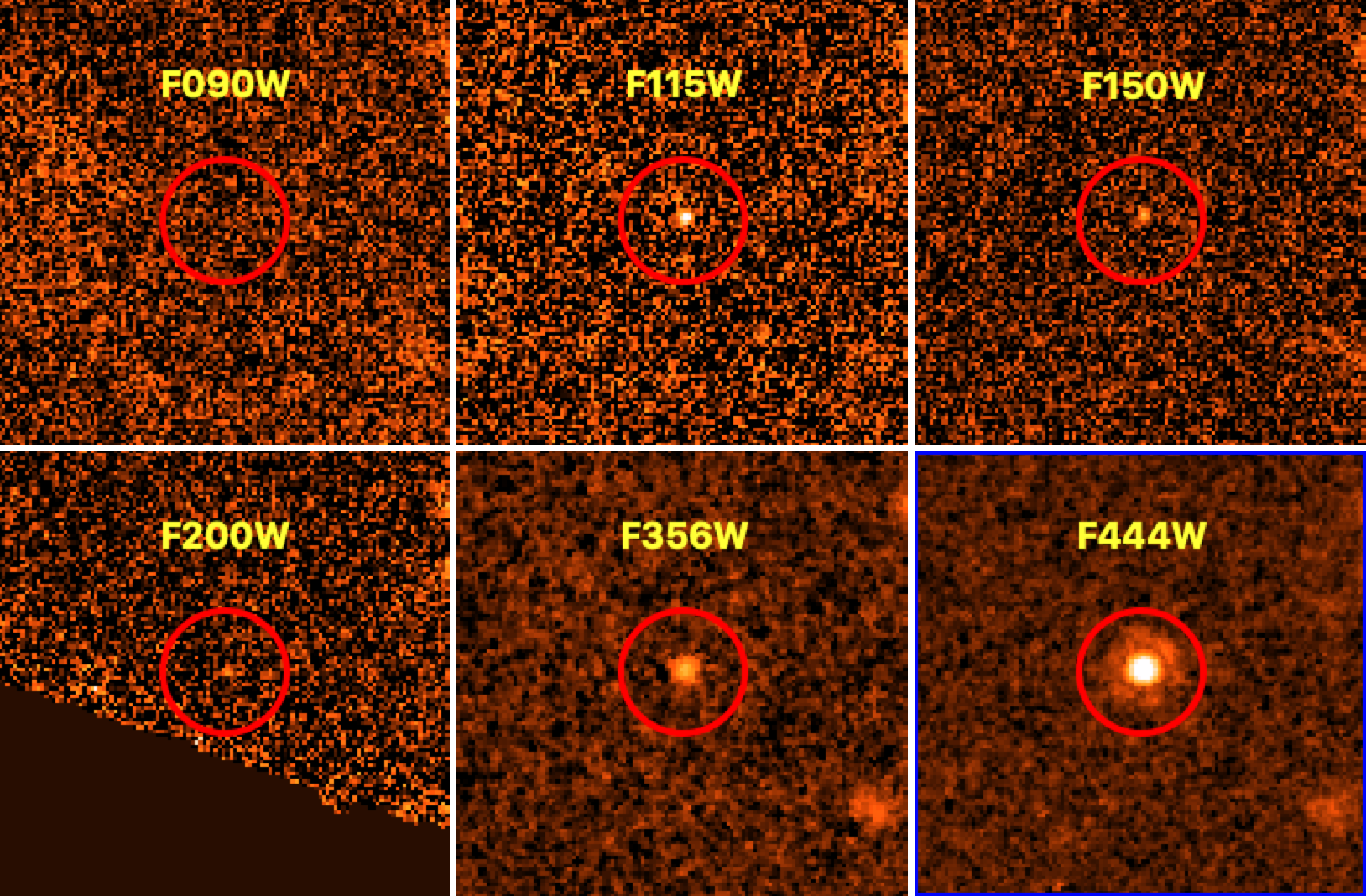}
\caption{Image cutouts (3$\farcs$5$\times$3$\farcs$5) of the GLASS-JWST-BD1 brown dwarf candidate.
The  circles have a radius of 0$\farcs$5. North is up, East is left.
\label{fig:1}}
\end{figure*}

In the NIRCam parallel data, we visually identified an isolated, very red, and unresolved source, which was particularly bright in F444W.
 This object immediately appeared to be a potential cool brown dwarf. 
To search for other cool brown dwarf candidates, we selected
from the extracted catalog of \citet{merlin2022} all sources with CLASS\_STAR $\geq$ 0.9 (point sources), $flags=0$, F444W $\leq$ 28 \citep[for reliable photometry in the other bands, see also][]{marco2022}, F356W$-$F444 $\geq$ 1.5, and F115W$-$F444W $\geq$ 2. 
The color selection criteria were motivated by the infrared colors of T and Y dwarfs,  specifically  Spitzer IRAC CH1$-$CH2 and J$-$CH2 colors \citep{meisner2020}. Only two sources passed these criteria, with one identified as a diffraction spike from a nearby bright star. We were thus left with  the original visually selected candidate, hereafter dubbed GLASS-JWST-BD1 ($\alpha$, $\delta$ = 00:14:03.33, -30:21:21.7). As an isolated source, we can rule out flux contamination from other sources as an explanation for its unusual colors. Figure~\ref{fig:1} displays images of the source in six filters, while Table~\ref{tab:mag} summarizes the relevant photometric data. The source is undetected in F090W.
We note that this location is covered by previous  Spitzer IRAC CH1 and CH2 observations (Hubble Frontier Fields Program, \citealt{lotz2017}), but the object is
significantly fainter than the CH2 $5\sigma$ limit of 24 in those observations \citep{sun2021}.


\begin{deluxetable*}{cccc}
\tablenum{1}
\tablecaption{Photometric data for GLASS-JWST-BD1. \label{tab:mag}}
\tablewidth{0pt}
\tablehead{
\colhead{\hspace{.0cm}Filter} &
\colhead{\hspace{2.0cm}AB} &
\colhead{\hspace{2.0cm}Vega} &
}
\startdata
F090W &\hspace{2cm} $\geq 29.0$ &\hspace{2cm} $\geq 28.5$ \\
F115W &\hspace{2cm} $27.766\pm0.122$ &\hspace{2cm} 27.006  \\
F150W &\hspace{2cm} $28.501\pm0.301$ &\hspace{2cm} 27.301  \\
F200W &\hspace{2cm} $29.530\pm1.345$ &\hspace{2cm} 27.680  \\
F277W &\hspace{2cm} $28.789\pm0.242$ &\hspace{2cm} 26.519  \\
F356W &\hspace{2cm} $27.572\pm0.073$ &\hspace{2cm} 24.812  \\
F444W &\hspace{2cm} $25.777\pm0.011$ &\hspace{2cm} 22.587 \\
\enddata
\tablecomments{
Magnitudes have been computed within a 0$\farcs$56 diameter aperture. AB to Vega conversion follows \cite{willmer2018}. }


\end{deluxetable*}

\section{analysis}
The near- and mid-infrared colors of low-temperature brown dwarfs are greatly affected by molecular absorption bands from H$_2$, H$_2$O, CH$_4$, CO$_2$, and NH$_3$ that dominate
their spectra \citep{2003ApJ...596..587B}. 
The Vega colors of  
GLASS-JWST-BD1, F115W$-$F444W = 4.55, and F356W$-$F444W = 2.26, are compatible with it being a late-type {\em T} dwarf, close to the {\em T/Y} boundary \citep{meisner2020,kirkpatrick2021}. To exploit the full photometric information from the NIRCam images, we performed a seven-band 
spectral energy distribution (SED) fit using  JWST/NIRCAM photometry provided with the  Sonora Cholla cloudless atmosphere models \citep{karalidi2021, marley2021}
\footnote{\href{https://zenodo.org/record/4450269\#.Ytqc-S8RrUo}{Models}}. We used the photometric table  \footnote{\href{https://zenodo.org/record/5063476/files/evolution_and_photometery.tar.gz}{Photometric table}} to compare models and observed photometry  in the full model parameter range, 200~K $\lesssim$ {\teff} $\lesssim$ 2400~K and 3 $\lesssim$ $\log{g}$ $\lesssim$ 5.5, provided in the solar metallicity table, with distance moduli spanning 5-15 magnitudes. The
 best-fit model of {\teff}=700~K and $\log{g}$=5.25 has a reduced $\chi^2$ = 6.65 if we include F277W data. However, if excluding this band, the 
best-fit model of {\teff}=650~K and $\log{g}=5.25$ has a smaller reduced $\chi^2$ = 0.80. 
The flux values of these two best-fit models differ by $<10\%$, and use the {\teff}=650~K model for our subsequent analysis.
The best scale factor between model and observed data that depend on the ratio of radius to distance as $\left(\frac{R}{d}\right)^2$, corresponds to a distance modulus of nine (630 pc)
 assuming $R$ = 0.83~R$_J$ as provided by the Sonora Cholla models (Fig 2).
This temperature corresponds to a spectral type of T8-T9 \citep{2019ApJS..240...19K}, consistent with its colors; and the temperature and surface gravity together correspond to a mass of $\sim$0.03~M$_{\odot}$ and an age of $\sim$5~Gyr based on the evolutionary models of \citet{marley2021}.
GLASS-JWST-BD1 is thus a substellar object.
We note that the Sonora-Cholla models use equilibrium chemistry, whereas
nonequilibrium chemistry is known to play an important  role in modulating molecular abundances in T dwarfs, notably  CO/CH$_4$ and NH$_3$/N$_2$ \citep{2006ApJ...647..552S,marley2015,leggett2021}. This may explain the disagreement in F277W photometry, and may also bias our model-inferred physical parameters. 
We consider these parameters preliminary estimates that may be improved with future modeling.


To determine the absolute magnitude and distance of this source, we used the magnitude/spectral type relations of \citet{kirkpatrick2021}.
Here, care needs to be taken with regard to filter systems
due to the complex spectral energy distributions of cool brown dwarfs. NIRCAM F444W is a close match to IRAC CH2, and in this band the source is brightest. 
Using the Sonora-Cholla spectra with {\teff} =  500--750K and $\log{g} = $  3 -- 5, we find a color offset of $-0.2 \lesssim$ CH2-F444W $\lesssim -0.1$, and adopt a value of $-0.15$.
Applying the absolute CH2/spectral type relation, we estimate absolute  F444W magnitudes of 13.65 (T8) and 14.16 (T9), corresponding to distance moduli of 9.30 (T8) and 8.79 (T9), similar to the model scale factor listed above.
We therefore infer a distance between 570~pc and 720~pc for GLASS-JWST-BD1, assuming it is a single T8--T9 dwarf.


The total survey volume for a T8--T9 dwarfs in the GLASS-ERS NIRCAM parallel field at our F444W = 28 search limit is approximately 7$\times$10$^4$~pc$^3$. 
\citet{kirkpatrick2021} measured a local space density of (5.4$\pm$0.6)$\times10^{-3}$~pc$^{-3}$ T8-T9 dwarfs, which would imply up to 370 such sources in the GLASS-ERS NIRCAM parallel field given a uniform spatial density. 
However, local brown dwarfs are largely members of the Galactic thin disk, and since
the Abell 2744 field is close to the South Galactic Pole 
({$l=8^{\circ}.9$, $b=-81^{\circ}$}), we are looking almost vertically out from the Galactic plane. 
Constraints on the thin disk vertical scale height for brown dwarfs remain uncertain.
\citet{ryan2011} report a scale height of 350~pc for late-M, L and T dwarfs in a deep {\it HST} imaging sample;
\citet{aganze2022} report a scale height of 175 pc -- 193 pc for T dwarfs with {\it HST} parallel spectroscopy. 
Based on these values, GLASS-JWST-BD1 is $\approx 2-5$ scale heights away from the Galactic plane.
Adopting a scale height of 300~pc and the local space density from \citet{kirkpatrick2021}, the predicted number of thin disk T8--T9 dwarfs in the Abell 2744 field would be closer to 0.24, which is unlikely, but not unreasonable.
Estimates for the space density or distribution of thick disk or halo brown dwarfs remain poorly constrained,
but the low number of predicted thin disk late T dwarfs in the Abell 2744 field suggests that GLASS-JWST-BD1 is likely a member of the thick disk or halo population, of which few examples are currently known \citep{meisner2020,2020ApJ...898...77S}.
%

We investigated other possible explanations for the SED of GLASS-JWST-BD1. 
The polar direction of this field means that Galactic extinction is negligible, making it highly unlikely that GLASS-JWST-BD1 is a reddened background star; this scenario is also ruled out by its relatively bright F115W magnitude.
We also attempted to fit a $z\sim 20$ galaxy model where the Lyman break
lies between F356W and F444W. This again cannot explain the emission near 1~{\micron} which would be completely absorbed by the neutral intergalactic medium; moreover, this solution predicts a much sharper drop between the F356W and F444W filters. We also tried to replicate the SED as a compact galaxy or low-luminosity quasar at $z\sim8.5$ with extreme [OIII]+H$\beta$ emission in F444W.  We fit the photometry to extragalactic templates from the {\em gsf} code \citep{morishita2019}. None of the templates using reasonable assumptions of line emission reproduce the observed data points better than the cool brown dwarf model
(minimum reduced $\chi^2 = 14.1$). 


\section{conclusions}
We report the serendipitous discovery of GLASS-JWST-BD1, a faint, distant, and low-temperature brown dwarf. Comparison with theoretical models suggests that the object is a late-type T dwarf with {\teff} $\approx$ 650~K.  
At its distance, the substellar population is likely dominated by thick disk and halo objects, and GLASS-JWST-BD1 may be such a source. 
To conclusively demonstrate this, 
kinematic or chemical abundance data are needed, 
requiring challenging spectroscopic observations even for {\it JWST}. Further information could be gleaned from narrowband imaging or longer-wavelength observations. 
The large estimated distance of GLASS-JWST-BD1 confirms the power of JWST to probe the very low-mass end of the stellar and substellar mass function in the Galactic thick disk and halo, enabling exploration of metallicity dependence on low-mass star formation and the evolution of brown dwarf atmospheres.





\section{Acknowledgments}
    This work is based on observations made with the NASA/ESA/CSA James Webb Space Telescope. The data were obtained from the Mikulski Archive for Space Telescopes at the Space Telescope Science Institute, which is operated by the Association of Universities for Research in Astronomy, Inc., under NASA contract NAS 5-03127 for JWST. These observations are associated with program JWST-ERS-1324. We acknowledge financial support from NASA through grant JWST-ERS-1324. K.G. and T.N. acknowledge support from Australian Research Council Laureate Fellowship FL180100060. 
    We acknowledge financial support through grants PRIN-MIUR 2017WSCC32, PRIN-MIUR  2020SKSTHZ and INAF-Mainstreams 1.05.01.86.20.  M.B. acknowledges support by the Slovenian national research agency ARRS through grant N1-0238.

\vspace{5mm}

\facilities{ {\it JWST} (NIRCam imaging)}
\software{{\em gsf} \citep{morishita2019}, SPLAT \citep{2017ASInC..14....7B}, ds9 \citep{2000ascl.soft03002S}, {\bf R} \citep{rcore}}

\bigskip

\begin{figure*}
\centering
    
\includegraphics[scale=0.6,angle=0]{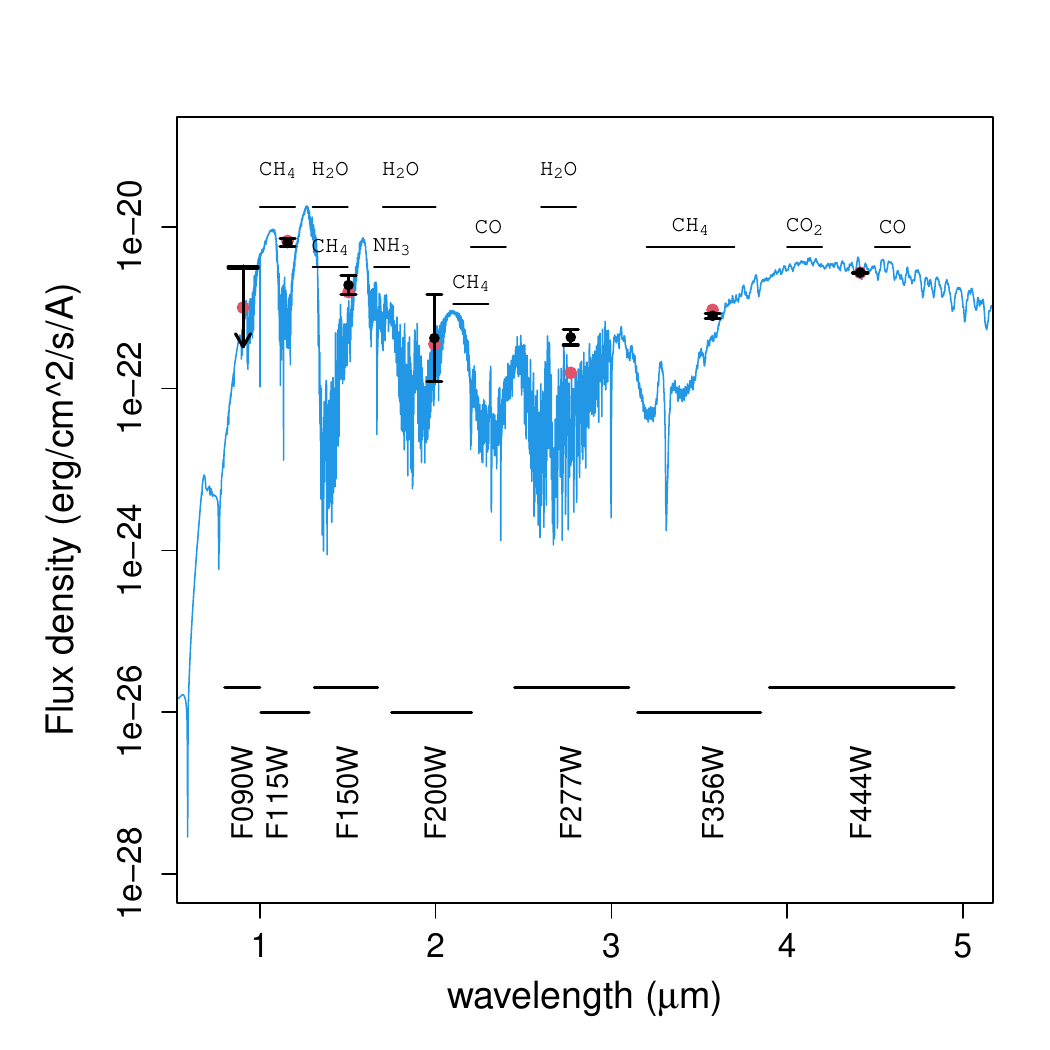}
\caption{Comparison of observed photometry (black points with 1$\sigma$ error bars) to the best-fit Sonora Cholla smoothed synthetic spectrum (blue line) and photometry (red points), with {\teff} = 650~K, $\log{g}$ = 5.25, and solar metallicity. We indicate the $>50\%$ range of the JWST NIRCam filter transmissions used in the GLASS-JWST-ERS observations.  Error bars are at $1\sigma$. Note that the source is not detected  in the  NIRCam F090W filter.
Most of the dominant infrared molecular absorption features of H2O, CO, CO2, CH4, and NH3 are shown.
}
\label{fig:2}
\end{figure*}

\bibliography{biblio.bib}{}

\begin{thebibliography}{}
\expandafter\ifx\csname natexlab\endcsname\relax\def\natexlab#1{#1}\fi
\providecommand{\url}[1]{\href{#1}{#1}}
\providecommand{\dodoi}[1]{doi:~\href{http://doi.org/#1}{\nolinkurl{#1}}}
\providecommand{\doeprint}[1]{\href{http://ascl.net/#1}{\nolinkurl{http://ascl.net/#1}}}
\providecommand{\doarXiv}[1]{\href{https://arxiv.org/abs/#1}{\nolinkurl{https://arxiv.org/abs/#1}}}

\bibitem[{{Aganze} {et~al.}(2022){Aganze}, {Burgasser}, {Malkan}, {Theissen},
  {Tejada Arevalo}, {Hsu}, {Bardalez Gagliuffi}, {Ryan}, \&
  {Holwerda}}]{aganze2022}
{Aganze}, C., {Burgasser}, A.~J., {Malkan}, M., {et~al.} 2022, \apj, 924, 114,
  \dodoi{10.3847/1538-4357/ac35ea}

\bibitem[{{Burgasser} \& {Splat Development Team}(2017)}]{2017ASInC..14....7B}
{Burgasser}, A.~J., \& {Splat Development Team}. 2017, in Astronomical Society
  of India Conference Series, Vol.~14, Astronomical Society of India Conference
  Series, 7--12.
\newblock \doarXiv{1707.00062}

\bibitem[{{Burgasser} {et~al.}(2003){Burgasser}, {Kirkpatrick}, {Burrows},
  {Liebert}, {Reid}, {Gizis}, {McGovern}, {Prato}, \&
  {McLean}}]{2003ApJ...592.1186B}
{Burgasser}, A.~J., {Kirkpatrick}, J.~D., {Burrows}, A., {et~al.} 2003, \apj,
  592, 1186, \dodoi{10.1086/375813}

\bibitem[{{Burrows} {et~al.}(2003){Burrows}, {Sudarsky}, \&
  {Lunine}}]{2003ApJ...596..587B}
{Burrows}, A., {Sudarsky}, D., \& {Lunine}, J.~I. 2003, \apj, 596, 587,
  \dodoi{10.1086/377709}

\bibitem[{{Castellano} {et~al.}(2022){Castellano}, {Fontana}, {Treu},
  {Santini}, {Merlin}, {Leethochawalit}, {Trenti}, {Mestric}, {Vanzella},
  {Bonchi}, {Belfiori}, {Nonino}, {Paris}, {Polenta}, {Roberts-Borsani},
  {Boyett}, {Calabro}, {Glazebrook}, {Grillo}, {Mascia}, {Mason}, {Mercurio},
  {Morishita}, {Nanayakkara}, {Pentericci}, {Rosati}, {Vulcani}, {Wang}, \&
  {Yang}}]{marco2022}
{Castellano}, M., {Fontana}, A., {Treu}, T., {et~al.} 2022, arXiv e-prints,
  arXiv:2207.09436.
\newblock \doarXiv{2207.09436}

\bibitem[{{Chabrier} \& {Baraffe}(2000)}]{2000ARA&A..38..337C}
{Chabrier}, G., \& {Baraffe}, I. 2000, \araa, 38, 337,
  \dodoi{10.1146/annurev.astro.38.1.337}

\bibitem[{{Karalidi} {et~al.}(2021){Karalidi}, {Marley}, {Fortney}, {Morley},
  {Saumon}, {Lupu}, {Visscher}, \& {Freedman}}]{karalidi2021}
{Karalidi}, T., {Marley}, M., {Fortney}, J.~J., {et~al.} 2021, \apj, 923, 269,
  \dodoi{10.3847/1538-4357/ac3140}

\bibitem[{{Kirkpatrick}(2005)}]{kirkpatrick2005}
{Kirkpatrick}, J.~D. 2005, \araa, 43, 195,
  \dodoi{10.1146/annurev.astro.42.053102.134017}

\bibitem[{{Kirkpatrick} {et~al.}(2019){Kirkpatrick}, {Martin}, {Smart},
  {Cayago}, {Beichman}, {Marocco}, {Gelino}, {Faherty}, {Cushing}, {Schneider},
  {Mace}, {Tinney}, {Wright}, {Lowrance}, {Ingalls}, {Vrba}, {Munn}, {Dahm}, \&
  {McLean}}]{2019ApJS..240...19K}
{Kirkpatrick}, J.~D., {Martin}, E.~C., {Smart}, R.~L., {et~al.} 2019, \apjs,
  240, 19, \dodoi{10.3847/1538-4365/aaf6af}

\bibitem[{{Kirkpatrick} {et~al.}(2021){Kirkpatrick}, {Gelino}, {Faherty},
  {Meisner}, {Caselden}, {Schneider}, {Marocco}, {Cayago}, {Smart},
  {Eisenhardt}, {Kuchner}, {Wright}, {Cushing}, {Allers}, {Bardalez Gagliuffi},
  {Burgasser}, {Gagn{\'e}}, {Logsdon}, {Martin}, {Ingalls}, {Lowrance},
  {Abrahams}, {Aganze}, {Gerasimov}, {Gonzales}, {Hsu}, {Kamraj}, {Kiman},
  {Rees}, {Theissen}, {Ammar}, {Andersen}, {Beaulieu}, {Colin}, {Elachi},
  {Goodman}, {Gramaize}, {Hamlet}, {Hong}, {Jonkeren}, {Khalil}, {Martin},
  {Pendrill}, {Pumphrey}, {Rothermich}, {Sainio}, {Stenner}, {Tanner},
  {Th{\'e}venot}, {Voloshin}, {Walla}, {W{\k{e}}dracki}, \& {Backyard Worlds:
  Planet 9 Collaboration}}]{kirkpatrick2021}
{Kirkpatrick}, J.~D., {Gelino}, C.~R., {Faherty}, J.~K., {et~al.} 2021, \apjs,
  253, 7, \dodoi{10.3847/1538-4365/abd107}

\bibitem[{{Kumar}(1962)}]{1962AJ.....67S.579K}
{Kumar}, S.~S. 1962, \aj, 67, 579, \dodoi{10.1086/108658}

\bibitem[{{Leggett} {et~al.}(2021){Leggett}, {Tremblin}, {Phillips}, {Dupuy},
  {Marley}, {Morley}, {Schneider}, {Caselden}, {Guillaume}, \&
  {Logsdon}}]{leggett2021}
{Leggett}, S.~K., {Tremblin}, P., {Phillips}, M.~W., {et~al.} 2021, \apj, 918,
  11, \dodoi{10.3847/1538-4357/ac0cfe}

\bibitem[{{Lotz} {et~al.}(2017){Lotz}, {Koekemoer}, {Coe}, {Grogin}, {Capak},
  {Mack}, {Anderson}, {Avila}, {Barker}, {Borncamp}, {Brammer}, {Durbin},
  {Gunning}, {Hilbert}, {Jenkner}, {Khandrika}, {Levay}, {Lucas}, {MacKenty},
  {Ogaz}, {Porterfield}, {Reid}, {Robberto}, {Royle}, {Smith},
  {Storrie-Lombardi}, {Sunnquist}, {Surace}, {Taylor}, {Williams}, {Bullock},
  {Dickinson}, {Finkelstein}, {Natarajan}, {Richard}, {Robertson}, {Tumlinson},
  {Zitrin}, {Flanagan}, {Sembach}, {Soifer}, \& {Mountain}}]{lotz2017}
{Lotz}, J.~M., {Koekemoer}, A., {Coe}, D., {et~al.} 2017, \apj, 837, 97,
  \dodoi{10.3847/1538-4357/837/1/97}

\bibitem[{Marley \& Robinson(2015)}]{marley2015}
Marley, M., \& Robinson, T. 2015, Annual Review of Astronomy and Astrophysics,
  53, 279, \dodoi{10.1146/annurev-astro-082214-122522}

\bibitem[{{Marley} {et~al.}(2021){Marley}, {Saumon}, {Visscher}, {Lupu},
  {Freedman}, {Morley}, {Fortney}, {Seay}, {Smith}, {Teal}, \&
  {Wang}}]{marley2021}
{Marley}, M.~S., {Saumon}, D., {Visscher}, C., {et~al.} 2021, \apj, 920, 85,
  \dodoi{10.3847/1538-4357/ac141d}

\bibitem[{{Masters} {et~al.}(2012){Masters}, {McCarthy}, {Burgasser}, {Hathi},
  {Malkan}, {Ross}, {Siana}, {Scarlata}, {Henry}, {Colbert}, {Atek},
  {Rafelski}, {Teplitz}, {Bunker}, \& {Dressler}}]{masters2012}
{Masters}, D., {McCarthy}, P., {Burgasser}, A.~J., {et~al.} 2012, \apjl, 752,
  L14, \dodoi{10.1088/2041-8205/752/1/L14}

\bibitem[{{Meisner} {et~al.}(2020){Meisner}, {Faherty}, {Kirkpatrick},
  {Schneider}, {Caselden}, {Gagn{\'e}}, {Kuchner}, {Burgasser}, {Casewell},
  {Debes}, {Artigau}, {Bardalez Gagliuffi}, {Logsdon}, {Kiman}, {Allers},
  {Hsu}, {Wisniewski}, {Allen}, {Beaulieu}, {Colin}, {Durantini Luca},
  {Goodman}, {Gramaize}, {Hamlet}, {Hinckley}, {Kiwy}, {Martin}, {Pendrill},
  {Rothermich}, {Sainio}, {Sch{\"u}mann}, {Andersen}, {Tanner}, {Thakur},
  {Th{\'e}venot}, {Walla}, {W{\k{e}}dracki}, {Aganze}, {Gerasimov}, {Theissen},
  \& {Backyard Worlds: Planet 9 Collaboration}}]{meisner2020}
{Meisner}, A.~M., {Faherty}, J.~K., {Kirkpatrick}, J.~D., {et~al.} 2020, \apj,
  899, 123, \dodoi{10.3847/1538-4357/aba633}

\bibitem[{{Merlin} {et~al.}(2022){Merlin}, {Bonchi}, {Paris}, {Belfiori},
  {Fontana}, {Castellano}, {Nonino}, {Polenta}, {Santini}, {Yang},
  {Glazebrook}, {Treu}, {Roberts-Borsani}, {Trenti}, {Birrer}, {Brammer},
  {Grillo}, {Calabr{\`o}}, {Marchesini}, {Mason}, {Mercurio}, {Morishita},
  {Strait}, {Boyett}, {Leethochawalit}, {Nanayakkara}, {Vulcani}, {Bradac}, \&
  {Wang}}]{merlin2022}
{Merlin}, E., {Bonchi}, A., {Paris}, D., {et~al.} 2022, arXiv e-prints,
  arXiv:2207.11701.
\newblock \doarXiv{2207.11701}

\bibitem[{{Morishita} {et~al.}(2019){Morishita}, {Abramson}, {Treu}, {Brammer},
  {Jones}, {Kelly}, {Stiavelli}, {Trenti}, {Vulcani}, \&
  {Wang}}]{morishita2019}
{Morishita}, T., {Abramson}, L.~E., {Treu}, T., {et~al.} 2019, \apj, 877, 141,
  \dodoi{10.3847/1538-4357/ab1d53}

\bibitem[{{R Core Team}(2020)}]{rcore}
{R Core Team}. 2020, R: A Language and Environment for Statistical Computing, R
  Foundation for Statistical Computing, Vienna, Austria.
\newblock \url{https://www.R-project.org/}

\bibitem[{{Ryan} {et~al.}(2011){Ryan}, {Thorman}, {Yan}, {Fan}, {Yan},
  {Mechtley}, {Hathi}, {Cohen}, {Windhorst}, {McCarthy}, \&
  {Wittman}}]{ryan2011}
{Ryan}, R.~E., {Thorman}, P.~A., {Yan}, H., {et~al.} 2011, \apj, 739, 83,
  \dodoi{10.1088/0004-637X/739/2/83}

\bibitem[{{Saumon} {et~al.}(2006){Saumon}, {Marley}, {Cushing}, {Leggett},
  {Roellig}, {Lodders}, \& {Freedman}}]{2006ApJ...647..552S}
{Saumon}, D., {Marley}, M.~S., {Cushing}, M.~C., {et~al.} 2006, \apj, 647, 552,
  \dodoi{10.1086/505419}

\bibitem[{{Schneider} {et~al.}(2020){Schneider}, {Burgasser}, {Gerasimov},
  {Marocco}, {Gagn{\'e}}, {Goodman}, {Beaulieu}, {Pendrill}, {Rothermich},
  {Sainio}, {Kuchner}, {Caselden}, {Meisner}, {Faherty}, {Mamajek}, {Hsu},
  {Greco}, {Cushing}, {Kirkpatrick}, {Bardalez-Gagliuffi}, {Logsdon}, {Allers},
  {Debes}, \& {Backyard Worlds: Planet 9 Collaboration}}]{2020ApJ...898...77S}
{Schneider}, A.~C., {Burgasser}, A.~J., {Gerasimov}, R., {et~al.} 2020, \apj,
  898, 77, \dodoi{10.3847/1538-4357/ab9a40}

\bibitem[{{Smithsonian Astrophysical Observatory}(2000)}]{2000ascl.soft03002S}
{Smithsonian Astrophysical Observatory}. 2000, {SAOImage DS9: A utility for
  displaying astronomical images in the X11 window environment}, Astrophysics
  Source Code Library, record ascl:0003.002.
\newblock \doeprint{0003.002}

\bibitem[{{Sun} {et~al.}(2021){Sun}, {Egami}, {P{\'e}rez-Gonz{\'a}lez},
  {Smail}, {Caputi}, {Bauer}, {Rawle}, {Fujimoto}, {Kohno},
  {Dudzevi{\v{c}}i{\={u}}t{\.{e}}}, {Atek}, {Bianconi}, {Chapman}, {Combes},
  {Jauzac}, {Jolly}, {Koekemoer}, {Magdis}, {Rodighiero}, {Rujopakarn},
  {Schaerer}, {Steinhardt}, {Van der Werf}, {Walth}, \& {Weaver}}]{sun2021}
{Sun}, F., {Egami}, E., {P{\'e}rez-Gonz{\'a}lez}, P.~G., {et~al.} 2021, \apj,
  922, 114, \dodoi{10.3847/1538-4357/ac2578}

\bibitem[{{Treu} {et~al.}(2022){Treu}, {Roberts-Borsani}, {Bradac}, {Brammer},
  {Fontana}, {Henry}, {Mason}, {Morishita}, {Pentericci}, {Wang}, {Acebron},
  {Bagley}, {Bergamini}, {Belfiori}, {Bonchi}, {Boyett}, {Boutsia}, {Calabro},
  {Caminha}, {Castellano}, {Dressler}, {Glazebrook}, {Grillo}, {Jacobs},
  {Jones}, {Kelly}, {Leethochawalit}, {Malkan}, {Marchesini}, {Mascia},
  {Mercurio}, {Merlin}, {Nanayakkara}, {Nonino}, {Paris}, {Poggianti},
  {Rosati}, {Santini}, {Scarlata}, {Shipley}, {Strait}, {Trenti}, {Tubthong},
  {Vanzella}, {Vulcani}, \& {Yang}}]{Treu2022}
{Treu}, T., {Roberts-Borsani}, G., {Bradac}, M., {et~al.} 2022, ApJ, in press,
  arXiv:2206.07978.
\newblock \doarXiv{2206.07978}

\bibitem[{{Willmer}(2018)}]{willmer2018}
{Willmer}, C. N.~A. 2018, \apjs, 236, 47, \dodoi{10.3847/1538-4365/aabfdf}

\bibitem[{{Zhang} {et~al.}(2019){Zhang}, {Burgasser}, {G{\'a}lvez-Ortiz},
  {Lodieu}, {Zapatero Osorio}, {Pinfield}, \& {Allard}}]{2019MNRAS.486.1260Z}
{Zhang}, Z.~H., {Burgasser}, A.~J., {G{\'a}lvez-Ortiz}, M.~C., {et~al.} 2019,
  \mnras, 486, 1260, \dodoi{10.1093/mnras/stz777}

\end{thebibliography}
\bibliographystyle{aasjournal}

\end{document}